\documentstyle[amsfonts]{article}
\input epsf

\newcommand{\bea}{\begin{eqnarray}}
\newcommand{\eea}{\end{eqnarray}}
\newcommand{\beq}{\begin{equation}}
\newcommand{\eeq}{\end{equation}}
\newcommand{\cd}{\partial}
\newcommand{\CP}{{\Bbb C}P^{1}}
\newcommand{\R}{{\Bbb R}}

\newcommand{\Z}{{\Bbb Z}}
\newcommand{\ra}{\rightarrow}
\newcommand{\vs}{\vspace{0.5cm}}
\newcommand{\Ibar}{\overline{I}_{ij}}
\newcommand{\wt}{\widetilde}

\title{Topological discrete kinks}
\author{J. M. Speight \\ Max-Planck-Institute f\"{u}r Mathematik in
den Naturwissenschaften \\ Inselstra{\ss}e 22-26, 04103 Leipzig, Germany}
\date{}

\begin{document}
\maketitle
\begin{abstract}
A spatially discrete version of the general kink-bearing nonlinear
Klein-Gordon model in $(1+1)$ dimensions is constructed which preserves the
topological lower bound on kink energy. It is proved that, provided the
lattice spacing $h$ is sufficiently small, there exist static kink solutions
attaining this lower bound centred anywhere relative to the spatial lattice.
Hence there is no Peierls-Nabarro barrier impeding the propagation of kinks
in this discrete system. An upper bound on $h$ is derived and given a physical
interpretation in terms of the radiation of the system. The construction,
which works most naturally when the nonlinear Klein-Gordon model has a
squared polynomial interaction potential, is applied to a recently proposed
continuum model of polymer twistons. Numerical simulations are presented
which demonstrate that kink pinning is eliminated, and radiative kink
deceleration greatly reduced in comparison with the conventional discrete
system. So even on a very coarse lattice, kinks behave much as they do in the
continuum. It is argued, therefore, that the construction provides a natural
means of numerically simulating kink dynamics in nonlinear Klein-Gordon models
of this type. The construction is compared with the inverse method of Flach,
Zolotaryuk and Kladko. Using the latter method, alternative spatial
discretizations of the twiston and sine-Gordon models are obtained which are
also free of the Peierls-Nabarro barrier.
\end{abstract}
\section{Introduction}
\label{sec:intro}

A major difficulty in the study of soliton dynamics in the context of high
energy physics is that the requirement of Lorentz invariance appears 
incompatible with integrability, at least
 for Lagrangian field theories in spacetimes of
realistic dimension. The PDEs of most interest, therefore, are nonintegrable,
and there is no prospect of finding interesting exact solutions of multisoliton
initial value problems. This is why numerical simulation is a popular and 
important tool. In performing numerical simulations one is forced to 
discretize space in some way, and this inevitably introduces fictitious 
discretization effects, which one should seek to minimize. 

The standard discretization of a field equation (in which spatial partial
derivatives are replaced by simple difference operators, nothing else being 
changed) replaces it with an infinite system of coupled ODEs, representing 
a network of identical oscillators, nearest neighbours being coupled by
springs. Such networks have been extensively studied in recent years
\cite{pey,boe,dun}, primarily because of their condensed matter and biophysical
applications. The crucial discretization effect encountered is that static
solutions may no longer be centred at an arbitrary position in space, but 
must lie exactly on a lattice site, or at the centre of a lattice cell. These
two types of static solution have different energies (one is a saddle point, 
the other a minimum), so there is an energy barrier, called the 
Peierls-Nabarro (PN) barrier, 
resisting the free propagation of solitons form cell to 
cell.
As a soliton moves through the lattice, its motion in and out of the PN 
potential excites small amplitude traveling waves in its wake
(radiation, or phonons) which drain its kinetic energy, 
causing it to decelerate. Sometimes
the kink slows down so much that it has insufficient energy to surmount the PN
barrier, whereupon it becomes pinned to a lattice cell. 

Clearly, this behaviour is very different from soliton dynamics in the
continuum, so if the standard discretization is to be 
used the height of the PN 
barrier must be made negligible by using a very fine spatial lattice, with
spacing $h$ of order (soliton width)/20, say. This is computationally 
expensive.
An alternative approach, due to Richard Ward, is to exploit the non-uniqueness
of the discretization process to find a discretization with no PN barrier at
all. The hope would then be that the discrete soliton dynamics would closely
mimic its continuum counterpart even on very coarse lattices, with say
$h$ of order soliton width. 

How does one find such a discretization? Ward's idea is to construct a discrete
system which naturally preserves the ``Bogomol'nyi'' properties of its
continuum counterpart, that is, a topological lower bound on soliton energy
which is attained
 by solutions of some sort of first order self-duality equation. Of
course, the continuum system one starts with must be of Bogomol'nyi type,
but most systems of interest in particle theory are (e.g.\ BPS monopoles,
instantons, sigma model lumps, vortices, kinks). 
The idea has proved very successful in the 
cases of sine-Gordon and $\phi^4$ kink dynamics \cite{war,spe}; that is,
it was found that kink pinning was eliminated, and radiative deceleration 
greatly reduced in comparison with the standard discretizations. Unfortunately,
work on higher dimensional systems (where the benefit of a coarse lattice
would be greatest) has been less encouraging \cite{war2}. It is possible to 
find discretizations of some planar models which preserve the Bogomol'nyi
bound on soliton energy (and this ensures that the solitons are more stable
than they would otherwise be), but the bound is unattainable, and there is
no discrete self-duality equation. So the PN barrier persists, although it is
reduced to some extent. Alternatively, by imposing rotational
symmetry one can find discretizations with saturable Bogomol'nyi bounds
\cite{lee,the} but then, of course, the system is effectively one-dimensional.

The purpose of this paper is to demonstrate that Ward's idea works completely
generally for relativistic one-dimensional kinks. More precisely, we will
construct a ``topological'' discretization of a general nonlinear 
Klein-Gordon theory in $\R^{1+1}$. This will, by construction, preserve the
familiar Bogomol'nyi bound on kink energy. We will then prove that there
exist static kinks saturating this bound centred {\em anywhere} relative
to the lattice, so that there is no PN barrier in this discretization. The
construction is a generalization of the topological discrete $\phi^4$ system
previously mentioned \cite{spe}. As an extra example, we apply the construction
to a recently proposed continuum model of polymer twistons \cite{baz}, and 
again find that elimination of the PN barrier leads to continuum-like dynamics
deep within the discrete regime.

\section{Relativistic kinks}
\label{sec:relkinks}

The nonlinear Klein-Gordon system consists of a scalar field
$\phi:\R^{1+1}\ra\R$ whose dynamics is governed by the Lagrangian $L=E_K-E_P$,
the kinetic and potential energy functionals being, respectively,
\bea
E_K&=&\frac{1}{2}\int_{-\infty}^{\infty}dx\, \dot{\phi}^2, \nonumber \\
E_P&=&\frac{1}{2}\int_{-\infty}^{\infty}dx\, \left[\left(\frac{\cd\phi}{\cd x}
\right)^2+F_c(\phi)^2\right].
\eea
The field equation is the Euler-Lagrange equation obtained by demanding that
$\phi$ be a local extremal of the action $S=\int dt\, L$. The following
restrictions are made on $F_c$ so as to ensure that kink solutions exist:
$F_c\in C^1(\R,\R)$,  $2\leq{\rm card}(F_c^{-1}(0))
\leq\aleph_0$, and for all 
$u\in F_c^{-1}(0)$, $F_c'(u)\neq0$. So the model has at least $2$ and at 
most countably
many degenerate zero vacua, which we label so that $F_c^{-1}(0)=\{u_i:i\in J\}$
where $J\subseteq\Z$ is a (perhaps unbounded) set of consecutive integers,
$0\in J$, and $i>j\Rightarrow u_i>u_j$. 
Given these restrictions, $F_c$ must change sign at every
$u_i$. It is convenient to label the vacua so that $F_c$ is positive
on $(u_i,u_{i+1})$ whenever $i$ is even (hence negative when $i$ is odd).
For any pair $i,j\in J$, $|i-j|=1$,
by a type $(i,j)$ kink we mean any finite energy
solution of the field equation with the
boundary behaviour
\beq
\label{bcs}
\phi(x,t)\ra\left\{\begin{array}{cc}
u_i & x\ra-\infty \\
u_j & x\ra\infty,
\end{array}\right.
\eeq
for all $t$. It would be conventional to call these ``antikinks'' in the
case where $i$ is odd, but the distinction is rather cumbersome for our 
purposes, and will not be made.

Static type $(i,j)$ kinks may be obtained by a Bogomol'nyi argument \cite{bog}.
To this end, let 
\beq
G(\phi):=\int_0^\phi d\psi\, F_c(\phi).
\eeq
Clearly, $G\in C^2(\R,\R)$ and $G'=F_c$. Stable static solutions are local
minimals of $E_P[\phi]$. Assuming $\phi:\R\ra\R$ has boundary behaviour
(\ref{bcs}), we have,
\bea
E_P[\phi]&=&\frac{1}{2}\int_{-\infty}^\infty dx\, \left[\frac{d\phi}{dx}-
(-1)^iF_c(\phi)\right]^2+(-1)^i\int_{-\infty}^{\infty} dx\, \frac{d\phi}{dx}
F_c(\phi) \nonumber \\
&\geq&(-1)^i\int_{u_i}^{u_j}d\phi\, F_c(\phi) \nonumber \\
&=& |G(u_j)-G(u_i)|.
\eea
The factor $(-1)^i$ is included to ensure an optimal bound whether $i$ is
even or odd. So, assuming type $(i,j)$ boundary behaviour, $E_P\geq
 |G(u_j)-G(u_i)|$, with equality if and only if the Bogomol'nyi equation,
\beq
\label{cbog}
\frac{d\phi}{dx}=(-1)^iF_c(\phi),
\eeq
holds.
This first order ODE is readily reduced to quadratures. Let $I_{ij}$ be the
open interval between $u_i$ and $u_j$. Then, for each $\phi_0\in I_{ij}$, 
there exists a unique solution of (\ref{cbog}) with boundary behaviour
(\ref{bcs}) and $\phi(0)=\phi_0$. The set of such solutions constitutes a
translation orbit of any one of them, since for each fixed $x$, $\phi(x)$ is
a monotonic function of $\phi_0$. So we may parametrize the moduli space of
static $(i,j)$ kinks, $M_{ij}^c$, by $\phi_0\in I_{ij}$, just as well as by
kink position, defined for example by $b=\phi^{-1}(\frac{1}{2}(u_i+u_j))$. 
The former
parametrization generalizes more naturally to discrete systems than does the
latter.

For purposes of comparison with the discrete case, we remark that
the full field equation is
\beq
\label{ceqm}
\ddot{\phi}=\frac{\cd^2\phi}{\cd x^2}-F_c'(\phi)F_c(\phi).
\eeq
Note that the static field equation is a second order ODE, while the
Bogomol'nyi equation (\ref{cbog}) is first order.

\section{Topological discretization}
\label{sec:topdis}

From now on, $x$ takes values in a lattice of spacing $h>0$, so $x\in h\Z=
\{0,\pm h,\pm 2h,\ldots\}$. We introduce the notation $f_+$ and $f_-$ for 
forward and backward shifted versions of any function $f:h\Z\ra \R$, that is
$f_\pm(x):=f(x\pm h)$. Also, we denote by $\Delta$ the forward difference
operator: $\Delta f:=h^{-1}(f_+-f)$. Discretization of the nonlinear
Klein-Gordon system proceeds by defining discrete versions of $E_K$ and
$E_P$. These take the form
\bea
E_K&=&\frac{h}{2}\sum_{x\in h\Z}\dot{\phi}^2, \nonumber \\
E_P&=&\frac{h}{2}\sum_{x\in h\Z}\left[(\Delta\phi)^2+F^2\right],
\eea
where $F$ is a function of $\phi,\phi_+$ which will be defined below. This
will have the correct continuum limit provided that $\lim_{\phi_+\ra\phi}
F(\phi,\phi_+)=F_c(\phi)$. 

We shall choose $F$ as follows:
\beq
F(\phi,\phi_+)=\left\{\begin{array}{cc}
\displaystyle{\frac{G(\phi_+)-G(\phi)}{\phi_+-\phi}} & \phi\neq\phi_+ \\ & \\
F_c(\phi) & \phi=\phi_+.
\end{array}\right.
\eeq
This looks singular when $\phi=\phi_+$, but in fact $F:\R^2\ra\R$ is $C^1$ as
may be shown by using local compactness of $\R$ and Taylor's Theorem. (The 
practicalities of evaluating $F$ in a computer algorithm will be discussed in 
section \ref{sec:twist}.) The important point is that for all $\phi,\phi_+$,
\beq
(\Delta\phi)F=\Delta G,
\eeq
and this allows one to make a Bogomol'nyi argument analogous to that in
section \ref{sec:relkinks}. Namely, assuming that $\phi$ has type $(i,j)$
boundary behaviour,
\bea
E_P[\phi]&=&\frac{h}{2}\sum_{x\in h\Z}\left[\Delta\phi-(-1)^iF\right]^2
+(-1)^ih\sum_{x\in h\Z}(\Delta\phi)F \nonumber \\
&\geq&(-1)^ih\sum_{x\in h\Z}\Delta G \nonumber \\
&=& |G(u_j)-G(u_i)|,
\eea
so this gives a lower bound on type $(i,j)$ kink energy, which is attained
if and only if
\beq
\label{dbog}
\Delta\phi=(-1)^iF,
\eeq
which will henceforth be called the discrete Bogomol'nyi equation (DBE).

What this argument shows is that {\em if}\,
 solutions of the DBE exist with the
correct boundary behaviour, then they are minimals of $E_P$ within their class,
and hence static solutions of the model. Existence of such solutions centred 
{\em anywhere}\, relative to the lattice will be proved in section
\ref{sec:proof}. Since all such solutions have energy $ |G(u_j)-G(u_i)|$,
the kinks experience no PN barrier.

The discrete field equation is again the Euler-Lagrange equation for the
action $S=\int dt\, (E_K-E_P)$:
\beq
\label{deqm}
\ddot{\phi}=\frac{\phi_+-2\phi+\phi_-}{h^2}-F(\phi,\phi_+)F_1(\phi,\phi_+)
-F(\phi,\phi_-)F_1(\phi,\phi_-),
\eeq
where $F_1$ is the partial derivative of $F:\R\times\R\ra\R$ with respect to
the first factor, and symmetry of $F$ (under $\phi\leftrightarrow\phi_+$)
has been used.  The static field equation is a second order nonlinear
difference equation, in contrast with the DBE (\ref{dbog}) which is first
order. 

The system should be compared with the standard discrete nonlinear Klein-Gordon
system whose kinetic energy is the same, but whose potential energy is
\beq
E_P=\frac{h}{2}\left[(\Delta\phi)^2+F_c(\phi)^2\right].
\eeq
So the topological discretization differs only in that its ``substrate''
potential has been symmetrically distributed over pairs of nearest neighbours.
In the standard discrete system, the Bogomol'nyi argument is lost. The
discrete field equation is
\beq
\label{ndeqm}
\ddot{\phi}=\frac{\phi_+-2\phi+\phi_-}{h^2}-F_c(\phi)F_c'(\phi).
\eeq

\section{Continuity of the kink moduli space}
\label{sec:proof}

It is convenient to make the following definition: a type $(i,j)$ static kink
(where $|i-j|=1$, as before) is a monotonic mapping $\phi:h\Z\ra\R$ having
$(i,j)$ boundary behaviour (\ref{bcs}) and satisfying the DBE (\ref{dbog})
everywhere. Since the continuous translation symmetry of the continuum model
has been broken to symmetry under integer translations by $h$, there is no
reason to expect {\em a priori}, the existence of a continuous ``translation
orbit'' of static kink solutions. Indeed, in conventional discretizations,
the $(i,j)$
kink moduli space (that is, the set of type $(i,j)$ static kink solutions)
is $M_{ij}(h)=\frac{h}{2}\Z$, the PN barrier being the
difference in $E_P$ between ``odd'' and ``even'' members of this set. By 
contrast, we will prove that the topological discrete system defined in section
\ref{sec:topdis} has static kinks centred anywhere relative to the lattice,
so $M_{ij}(h)=\R$, at least for $h$ sufficiently small.

The idea behind the proof, which will be presented only for the case of even
$i$ (the odd $i$ case follows {\em mutatis mutandis}), is that any pair
$(\phi,\phi_+)$ satisfying the DBE (\ref{dbog}) may be associated with a point
on the intersection curve of the plane $\phi-\phi_++hz=0$ with the surface
$z=F(\phi,\phi_+)$ in $(\phi,\phi_+,z)$ space. If $h=0$, the plane is vertical,
and the vertical projection of this curve onto the $(\phi,\phi_+)$ plane is
clearly the graph of a function $g$ interpolating between $(u_i,u_i)$ and 
$(u_j,u_j)$ (namely, $g={\rm Id}$). If $h>0$, but very small, the plane tilts
slightly away from the vertical, but one may show that such a function 
persists, and the sequence generated by iteration of $g$ on any 
$\phi_0\in I_{ij}$,
a solution of the DBE by construction, has the required convergence properties.

\vs
\noindent
{\bf Theorem} For any pair $i,j\in J$ such that $|i-j|=1$, there exists
$h_*>0$ such that for all $h\in(0,h_*)$ and for all $\phi_0\in I_{ij}$ there
exists a type $(i,j)$ static kink with $\phi(0)=\phi_0$.
\vs

\noindent 
{\it Proof:} Let $i$ be even. 
Given any $C^1$ function $m:\R^n\ra\R$, $m_p$ denotes its partial derivative
with respect to the $p$-th factor. Also, any bounded open interval
$(z-\epsilon,z+\epsilon)$ will be denoted $B_\epsilon(z)$.
We first note that the only fixed points of the DBE
\beq
\label{dbe}
\frac{\phi_+-\phi}{h}=F(\phi,\phi_+)
\eeq
are the vacua, since $\phi_+=\phi\Rightarrow F(\phi,\phi)=F_c(\phi)=0$. Now,
one may regard a pair $(\phi,\phi_+)$ satisfying (\ref{dbe}) as a zero
of the $C^1$ function $f:\R^3\ra\R$
\beq
f(\phi_+,\phi,h)=\phi_+-\phi-hF(\phi,\phi_+).
\eeq
Note that for any $\psi\in\R$, $f(\psi,\psi,0)=0$ and 
$f_1(\psi,\psi,0)=1\neq 0$,
so by the Implicit Function Theorem, for every $\psi\in\R$ there exists
$\epsilon_\psi>0$ and a $C^1$ function 
$g:B_{\epsilon_\psi}(\psi)\times B_{\epsilon_\psi}(0)\ra\R$ 
such that $\phi_+=g_\psi(\phi,h)$ is the unique solution
of $f(\phi_+,\phi,h)=0$ in $B_{\epsilon_\psi}(\psi)$
for any $(\phi,h)\in{\rm dom}g_\psi$. Consider any
closed bounded interval $I\subset\R$. Since $I$ is compact, the open 
covering $\{B_{\epsilon_\psi}(\psi):\psi\in I\}$ has a finite
subcovering, $\{B_{\epsilon_n}(\psi_n):n=1,2,\ldots,N\}$, which
can be used to construct a $C^1$ function $g:I\times B_\epsilon(0)
\ra\R$ where $\epsilon=\inf_n\epsilon_n>0$. Namely,
\beq
g(\phi,h)=g_{\psi_n}(\phi,h)\qquad \mbox{if}\quad \phi\in
(\psi_n-\epsilon_n,\psi_n+\epsilon_n).
\eeq
This is well defined by the local uniqueness of each $g_{\psi_n}$. For each
$(\phi,h)\in I\times(-\epsilon,\epsilon)$, $\phi_+=g(\phi,h)$ is the locally
unique solution of (\ref{dbe}).

Consider the case $I=\Ibar$, the closure of the interval $I_{ij}$. We claim
that there exists $h_*\in(0,\epsilon)$ such that for every $h\in[0,h_*)$,
$g(\cdot,h):\Ibar\ra\R$ is strictly increasing. To see this, note that
$g(\cdot,0)={\rm Id}$, so 
$g_1(\phi,0)=1>0$. Now
$g_1$ is uniformly continuous on, say,
$\Ibar\times[-\frac{\epsilon}{2},\frac{\epsilon}{2}]$, so there exists
$h_*\in(0,\frac{\epsilon}{2}]$ such that for all $h\in[0,h_*)$
$g_1(\cdot,h):\Ibar\ra\R$ is
bounded away from $0$. 

Since the endpoints $u_i,u_j$ are fixed points of $g$, an immediate
consequence is that for all $h\in(0,h_*)$, $g(\cdot,h)$ is an increasing
$C^1$ mapping $\Ibar\ra\Ibar$, which we denote by simply $g$ for the rest
of this proof. Hence, for any $\phi_0\in I_{ij}$ the 
double-sided iteration sequence $\phi(nh)=g^n(\phi_0)$ 
exists and is a solution
of the DBE (\ref{dbe}) ($g^0:={\rm Id}$ and for $n<0$, $g^n
:=(g^{-1})^{|n|}$,
the existence of the inverse being guaranteed since $g$ is 
increasing). By monotonicity of $g$, for fixed $n\in\Z$, $\phi(nh)$ is
an increasing function of $\phi_0$. There can be no fixed points 
of $g$ other than the
endpoints (because $|i-j|=1$), so either
\bea
\mbox{case A:}\qquad g(\phi)&>&\phi\quad\forall\phi\in I_{ij} \nonumber \\
\mbox{or case B:}\qquad g(\phi)&<&\phi\quad\forall\phi\in I_{ij} \nonumber.
\eea
Given our definition of $F_c$, case A occurs when $i<j$ and case B when $i>j$
(recall we have assumed that $i$ is even).

Case A: the iteration sequence is increasing, bounded above by $u_j$ and below
by $u_i$ and hence both left and right convergent. Since $\Ibar$ is closed, it
contains $L,R$, the left and right limits. Consider the image of the forward
sequence $\phi(nh)$, $n\geq 1$ under $g$. Since $g$ is continuous,
$g(\phi(nh))\ra g(R)$. But $g(\phi(nh))$ is a subsequence of $\phi(nh)$ and
hence converges to $R$. Hence $R$ is a fixed point of $g$, and so $R=u_j$.
Similarly, considering the image of the sequence $\phi(-nh)$, $n\geq 1$, under
$g^{-1}$ one concludes that $L=u_i$. Hence $\phi$ has the correct limiting
behaviour.

Case B: now the iteration sequence is decreasing and bounded above by $u_i$ 
and below by $u_j$. A similar argument shows that $\phi(nh)\ra u_i$ as
$n\ra-\infty$ and $\phi(nh)\ra u_j$ as $n\ra\infty$. This completes the
proof. $\Box$
\vs

An important point to note is that $h_*$, the upper bound on lattice spacing
up to which $M_{ij}(h)$ is continuous (henceforth assumed to be optimal), 
depends in general on $(i,j)$. If $F_c^{-1}(0)$ is finite, a global upper
bound on $h$ may be found, namely,
\beq
h_*^g:=\inf\{h_*(i,j):i,j\in J, |i-j|=1\}>0
\eeq
so that the topological discretization with $0<h<h_*^g$ has continuous moduli
spaces for all types of kink. This is not necessarily true if $F_c^{-1}(0)$
is infinite (for then it is possible that $h_*^g=0$). For any particular
$(i,j)$, one can find an upper bound on $h_*$ however, meaning that in no
case does continuity of $M_{ij}(h)$ persist for all $h>0$.

The upper bound is obtained by again thinking of the graph of the function
$g(\cdot,h)$, whose existence was proved above, as the vertical projection of
the intersection of the plane $\phi-\phi_++(-1)^ihz=0$ with the surface
$z=F(\phi,\phi_+)$. Recall that this graph always passes through $(u_k,u_k)$,
where $k=i,j$. Consider the behaviour of the tangent to the graph at
$(u_k,u_k)$. As $h$ grows large, this tangent passes either through the 
horizontal or through the vertical (depending on whether $i$ is odd or even,
and whether $k=i$ or $k=j$). In either case, from this point on (i.e.\
for larger $h$) either monotonicity or existence of $g(\cdot,h)$ is lost,
and so the existence or correct limiting behaviour of the iteration sequence
$\phi(nh)=g^n(\phi_0,h)$ is generically lost. This will be explained in more
detail shortly.

The gradient of the tangent through $(u_k,u_k)$ may be calculated by demanding
that $(\phi,\phi_+)=(u_k+\delta,u_k+\delta_+)$ remains a solution of the
DBE (\ref{dbog}) to leading order
in $\|\delta\|=[\delta^2+\delta^2_+]^\frac{1}{2}$. Using
Taylor's Theorem for $G$,
\beq
G(u_k+\delta)=G(u_k)+\frac{\delta^2}{2}F_c'(u_k)+o(\delta^2),
\eeq
one finds that
\beq
\frac{\delta_+}{\delta}=\frac{2+(-1)^ihF_c'(u_k)}{2-(-1)^ihF_c'(u_k)}
+o(\|\delta\|).
\eeq
So if $i$ is even, meaning that $F_c'(u_i)>0$ and $F_c'(u_j)<0$, the tangent
through $(u_i,u_i)$ is vertical when $h=2/F_c'(u_i)$, and the tangent through
$(u_j,u_j)$ is horizontal when $h=2/|F_c'(u_j)|$. On the other hand, if $i$ is
odd, then $F_c'(u_i)<0$ and $F_c'(u_j)>0$, so the tangent through $(u_i,u_i)$
is horizontal when $h=2/|F_c'(u_i)|$, and the tangent through $(u_j,u_j)$
is vertical when $h=2/F_c'(u_j)$. In either case, we obtain the same upper
bound on $h_*$, namely
\beq
h_*\leq h_1 :=\inf\left\{\frac{2}{|F_c'(u_i)|},\frac{2}{|F_c'(u_j)|}\right\}.
\eeq
One may similarly define
\beq
h_2:=\sup\left\{\frac{2}{|F_c'(u_i)|},\frac{2}{|F_c'(u_j)|}\right\},
\eeq
which will prove useful below.

A typical situation is depicted graphically in figure 1. Here, 
$h_1=2/|F_c'(u_1)|$, $h_2=2/|F_c'(u_0)|$ and $h_1<h<h_2$, so the tangent
through $(u_1,u_1)$ has passed through the horizontal. It is impossible for
$\phi(nh)$ to converge {\em monotonically}\, to $u_1$ with $u_1$ as a cluster
point of the sequence: the sequence must overshoot $u_1$ as $\phi(nh)$
exceeds $\varphi_h$. The only possibility for a $(0,1)$ kink to exist for this
value of $h$, therefore, is that $\phi$ has a constant right hand tail, that
is, there exists $N\in\Z$ such that $\phi(nh)=u_1$ for all $n\geq N$. This, in
turn, is possible only if $\phi(0)=\varphi_h$, or one of its preimages under 
the iteration. One can restate this condition as $\phi(0)\in\Phi_h$ where
\beq
\Phi_h:=\{\gamma_h^n(\varphi_h):n=0,1,2,\ldots\},
\eeq
$\gamma_h$ being the partial inverse of $g(\cdot,h)$, with
$\gamma_h:[u_0,u_1]\ra[u_0,\varphi_h]$. The set $\Phi_h$ parametrizes a 
discrete moduli space of kinks $M_{01}(h)$, all identical modulo lattice
translations. Note that the left hand boundary behaviour of $\phi$
remains good for all $\phi(0)\in(u_0,u_1)$ in this situation.

\vspace{0.25cm}
\vbox{
\centerline{\epsfysize=2truein
\epsfbox{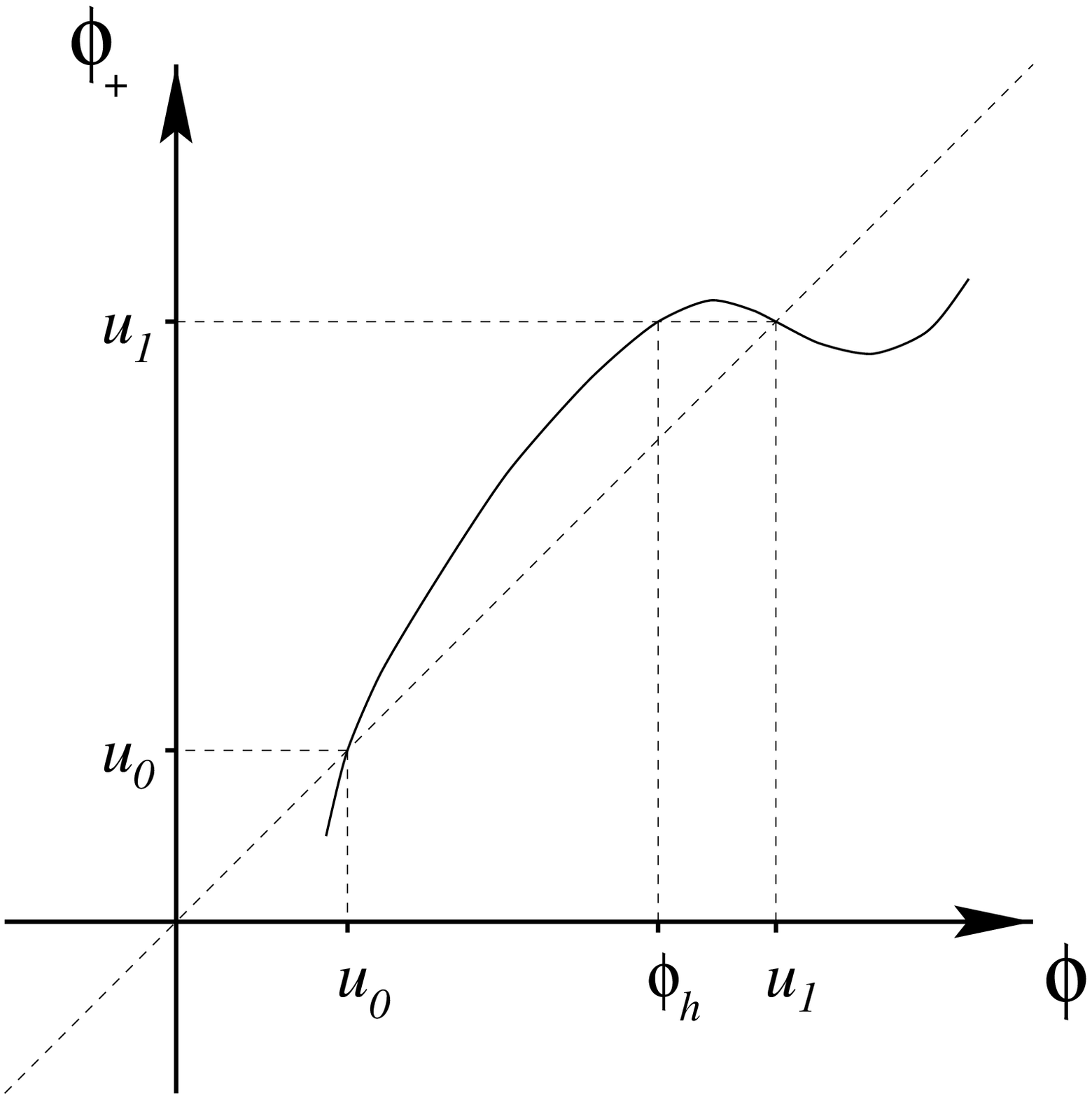}}
\begin{center}
{\it Figure 1: A section of the curve $\Delta\phi=F$, with $h_1<h<h_2$.}
\end{center}
}
\vspace{0.5cm}

In figure 2, $h$ has been increased further, so that $h>h_2$, and now the
sequence $\phi$ behaves badly at both ends. In order for $\phi(nh)\ra u_0$
monotonically as $n\ra-\infty$, $\phi$ must have a constant left tail, so one
must have $\phi(0)\in\wt{\Phi}_h$, where
\beq
\wt{\Phi}_h:=\{\wt{\gamma}_h^n(\wt{\varphi}_h):n=0,1,2,\ldots\},
\eeq
and $\wt{\gamma}_h:[u_0,\varphi_h]\ra[\wt{\varphi}_h,u_1]$ is defined by the
curve shown. In this case, 
for $\phi$ to satisfy the definition of a static kink, both
the left and right tails must be constant, which can only happen if
$\wt{\Phi}_h=\Phi_h$. One expects this to occur only for a discrete set of
$h$ values, which accumulate towards $h_2$. When $h$ becomes very large, one
expects the projected intersection curve to pass outside the open square
$(u_0,u_1)\times(u_0,u_1)$ completely, so that no static kink solutions are
possible. 

\vspace{0.25cm}
\vbox{
\centerline{\epsfysize=2truein
\epsfbox{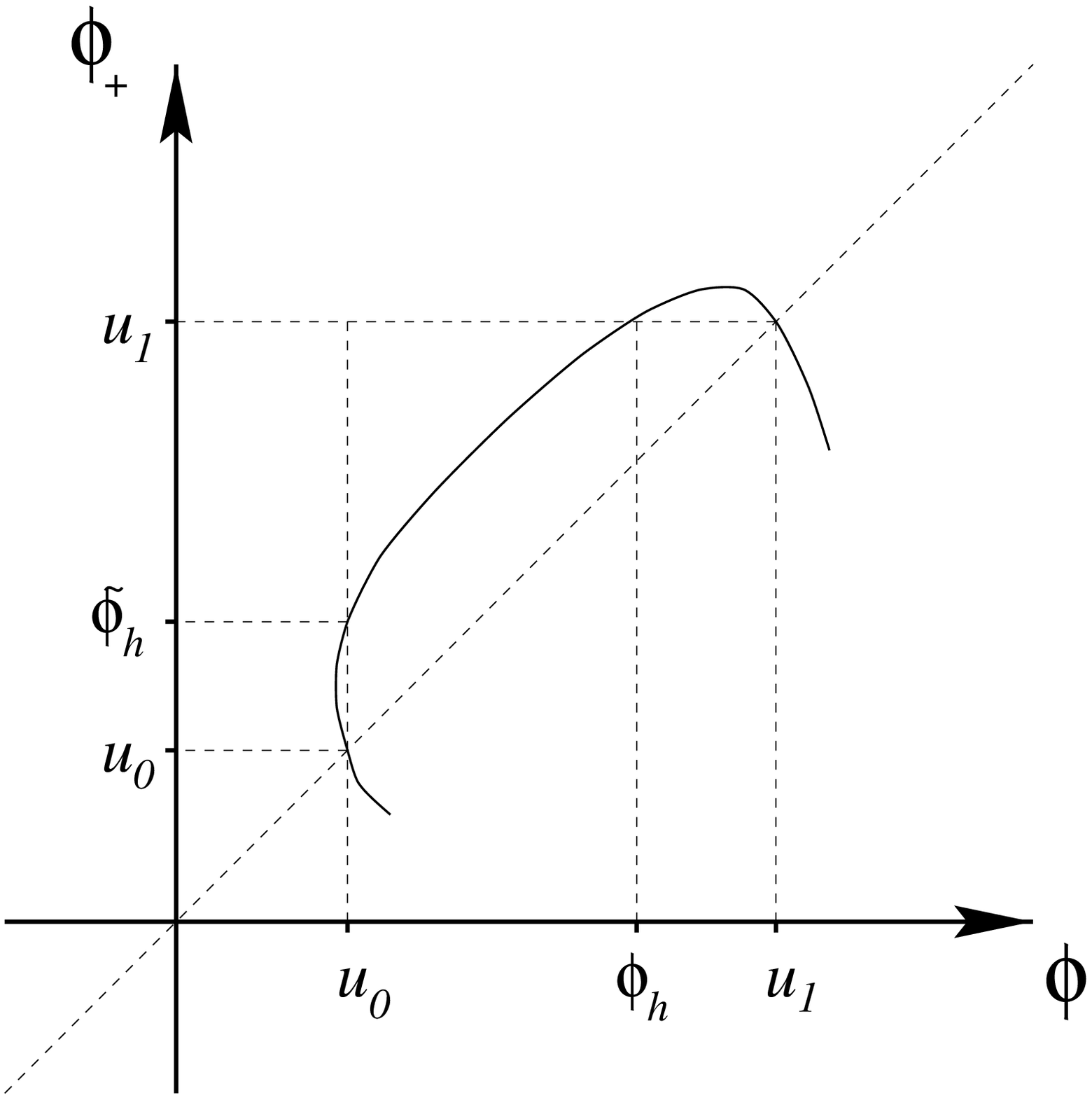}}
\begin{center}
{\it Figure 2: A section of the curve $\Delta\phi=F$, with $h>h_2$.}
\end{center}
}
\vspace{0.5cm}

We are led, therefore, to the following conjecture for the generic
behaviour of $M_{ij}(h)$ as $h$ varies: for $h\in(0,h_1]$, $M_{ij}(h)$ is
continuous, $\R$; for $h\in(h_1,h_2]$, $M_{ij}(h)$ is discrete, $h\Z$; for
almost all $h\in(h_2,\infty)$ $M_{ij}(h)=\emptyset$, but there exists a
bounded countable set $H\subset(h_2,\infty)$ with $\inf H=h_2$ such that
$M_{ij}(h)=h\Z$ for all $h\in H$. The conjecture is represented
schematically in figure 3.\, It matches the observed behaviour for the
topological discrete sine-Gordon \cite{war} and $\phi^4$ \cite{spe}
systems. In both these cases $h_1= h_2$ because the kinks interpolate
between identical vacua, so the middle band is missing, and for the
sine-Gordon model $H=\emptyset$. The example of twistons, discussed in
section \ref{sec:twist} produces the full generic behaviour, with all three
bands. It is probably possible to construct
a function $F_c$ such that this conjectured behaviour fails (by making
$F_c(\frac{1}{2}(u_i+u_j))$ very close to zero, for example), but such a system
is rather contrived. 
One should note that the onset of 
discreteness in the kink moduli space is due to the requirement that kinks
should be monotonic (after all, they are supposed to model continuum kinks,
which always are), rather than the appearance of some kind of
PN barrier.

\vspace{0.25cm}
\vbox{
\centerline{\epsfysize=2truein
\epsfbox{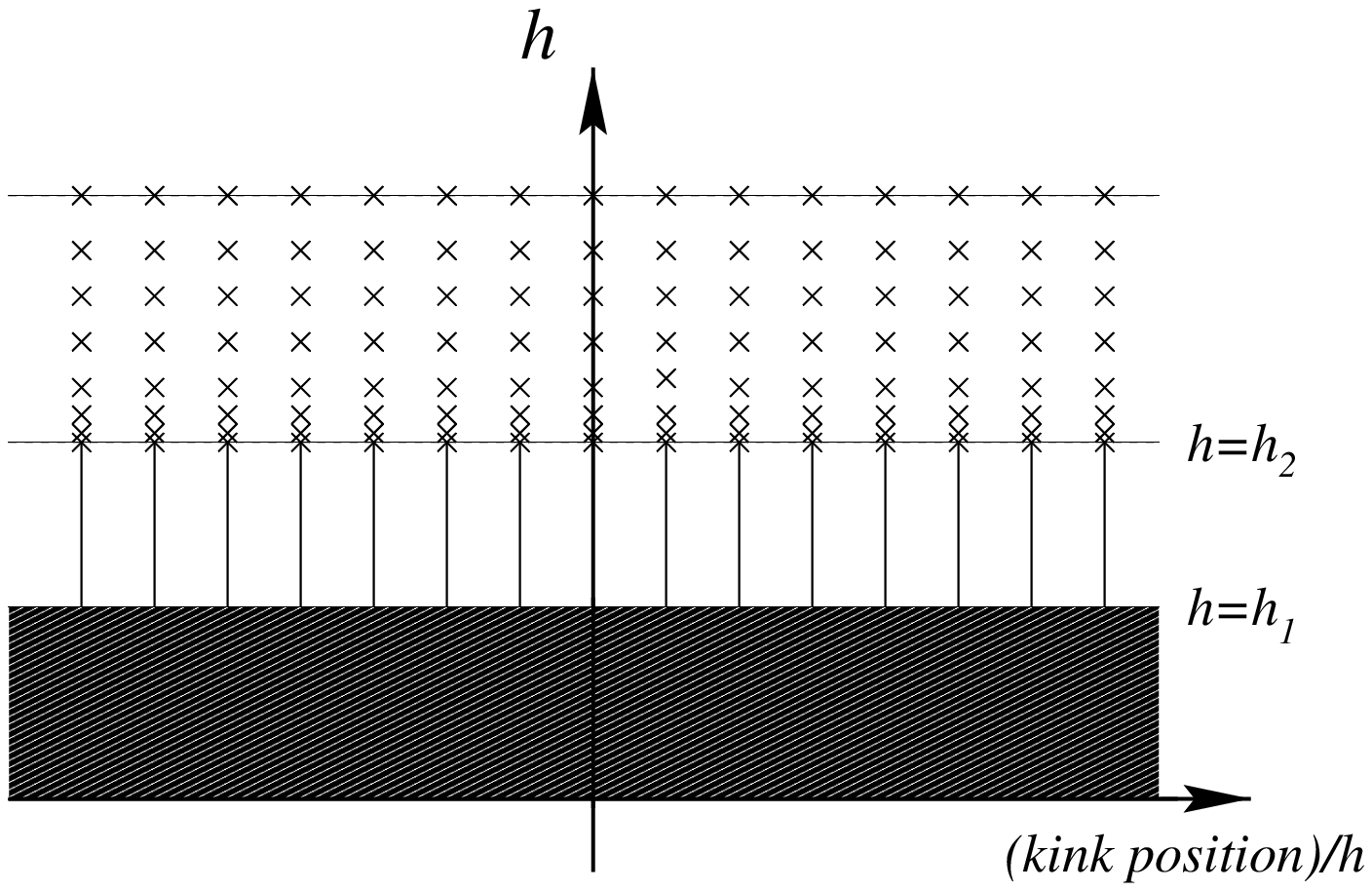}}
\noindent
{\it Figure 3: Schematic picture of the generic structure of the kink moduli
space $M_{ij}(h)$ as $h$ varies. Horizontal sections represent the moduli
spaces for each specific value of $h$.}
}
\vspace{0.5cm}

The upper bound $h_*\leq h_1$
has a remarkable physical interpretation in terms of the
radiation of the system. Just as different pairs of vacua $u_i,u_j$ have
different types of kink interpolating between them, so each different vacuum 
 $u_k$ has its own type of radiation. In each case, the dispersion relation
for this radiation may be obtained by substituting a traveling wave ansatz,
$\cos(kx-\omega t)$, into the discrete field equation linearized about $\phi=
u_k$. Noting that $F_1(u_k,u_k)=\frac{1}{2}F'_c(u_k)$, one finds that $\omega$
and $k$ must be related by
\beq
\omega^2=[F_c'(u_k)]^2+\left[\frac{4}{h^2}-(F_c'(u_k))^2\right]\sin^2\left(
\frac{kh}{2}\right).
\eeq
Note that this sinusoidal dispersion relation collapses to a flat line when
$h=2/|F_c'(u_k)|$, and that for larger $h$, the curve has a maximum instead
of a minimum at $k=0$. For $h\geq 2/|F_c'(u_k)|$, then, the discrete system
fails to model accurately the radiation of the continuum system (which obeys
the relativistic energy-momentum relation $\omega^2=[F_c'(u_k)]^2+k^2$) even
in the long wavelength limit. So the upper bound on $h_*$ derived above is
precisely that value of $h$ for which
 radiation about one or other of the vacua
between which the kinks interpolate starts to behave badly.  

\section{Twistons}
\label{sec:twist}

The original motivation for the construction in section \ref{sec:topdis} was
to provide an efficient and natural means of simulating soliton dynamics
numerically, on a computer. In this light, two substantial objections to
the discretization procedure can be raised. First, even if $F_c$ is known in
closed form, its anti-derivative $G$ may not be. In this case, every evaluation 
of $F$ in the computer program would require the approximate evaluation
of two definite integrals ($G(\phi)$ and $G(\phi_+)$). Clearly this cannot be
computationally efficient. Second, the piecewise definition of $F$ is highly
inconvenient for numerical purposes. That is, even if an explicit formula for
$G$ exists, evaluation of $F$ for $\phi$ close to $\phi_+$ would be subject to
significant rounding errors, unless (for example) a polynomial approximation
to $F$ were used in a narrow strip containing the diagonal $\phi=\phi_+$.
Again, this is complicated and inefficient. There is, however, a large class
of potentials $F_c^2$ for which both objections are avoided, namely those
where $F_c$ is polynomial. In this case, $G$ is also polynomial (hence
available in closed form) and, by the remainder theorem, so is
$[G(\phi_+)-G(\phi)]/(\phi_+-\phi)$, which may obviously be extended to the
whole $(\phi,\phi_+)$ plane to give a global, rather than piecewise,
definition of $F$. 

One example of such a polynomial discrete system has already been investigated
\cite{spe}. To illustrate the construction further, we shall investigate
the topological discretization of a recently proposed continuum model
of so-called twistons in crystalline polyethylene \cite{baz}. We emphasise
that the aim here is {\em not}\, to propose a physically realistic discrete
model of such polymer twistons, but rather to demonstrate that our discrete
system performs favourably in comparison with the conventional discretization
when simulating soliton dynamics in the {\em continuum}\, model.

The continuum system has interaction potential $F_c^2$, where
\beq
F_c(\phi)=\phi(1-\phi^2).
\eeq
We have normalized $F_c$ differently from \cite{baz}, and absorbed a coupling 
constant. So, there are three zero vacua, $u_0=-1$, $u_1=0$ and $u_2=1$, and
four different types of kink. Actually, all four types are trivially related
to one another by reflexions in space and/or the codomain, and so we shall
consider only type $(0,1)$ kinks. The static profiles of these may be found
explicitly using the Bogomol'nyi argument outlined in section 
\ref{sec:relkinks}, namely,
\beq
\phi(x)=-\sqrt{\frac{1}{2}(1-\tanh(x-b))}.
\eeq

Turning now to the topological discretization of this model (so now
$x\in h\Z$), we find that $G(\phi)=\phi^2(\phi^2-2)/4$, and
\beq
F(\phi,\phi_+)=\frac{G(\phi_+)-G(\phi)}{\phi+-\phi}=\frac{1}{4}
(\phi_+^2+\phi^2-2)(\phi_++\phi).
\eeq
As remarked above, $F$ has a simple closed form, and its evaluation presents
no problems for a computer program. By the general existence theorem of 
section \ref{sec:proof}, there exists $h_*>0$ such that for all $h\in(0,h_*)$
there is a continuum of $(0,1)$ kink solutions $M_{01}(h)$ of the DBE,
\beq
\label{twbog}
\frac{\phi_+-\phi}{h}=\frac{1}{4}(\phi_+^2+\phi^2-2)(\phi_++\phi),
\eeq
parametrized by $\phi(0)\in(-1,0)$. All such solutions have energy
$E_P=|G(0)-G(-1)|=\frac{1}{4}$, so there is no PN barrier. Recall that, in
general, one
has 
an upper bound $h_1$ on $h_*$. In the present case, $F_c'(-1)=-2$ and $F_c'(0)
=1$, so $h_1=1$ and $h_2=2$.

We do not propose to prove rigorously that $h_*=h_1=1$ 
for this model. Rather, by
plotting the zero contours of the function
\beq
f(\phi_+,\phi,h)=\phi_+-\phi-\frac{h}{4}(\phi_+^2+\phi^2-2)(\phi_++\phi)
\eeq
for various values of $h$ we will demonstrate that the generic
behaviour of $M_{ij}(h)$
described in section \ref{sec:proof} is highly plausible.
Recall, from the proof in that section that this contour contains,
at least for small $h$, the graph of the function $g(\cdot,h)$ which one
iterates to generate the static kink. Zero contours of $f$ for $h=0.5,1.0,1.5,
2.0,2.5$ and $4$, found using the contourplot utility of Maple, are shown in
figure 4. While $h<1$, the contour contains the graph of a continuous
increasing function $g(\cdot,h):[-1,0]\ra[-1,0]$, so that $M_{01}(h)$ is
continuous. However, when $h=1$, the tangent to the contour at $(-1,-1)$
passes through the vertical, so for $h>1$ only kinks with a constant left
tail can exist, which can only happen for nongeneric $\phi(0)$,
and $M_{ij}(h)$ is discrete. When $h=2$, the tangent to the
contour at $(0,0)$ passes through the horizontal. For $h>2$ kinks must also
have a constant right hand tail, so they may only exist for nongeneric 
$\phi(0)$ and $h$. This situation continues until $h\geq 4$, when the contour 
passes outside the open square $(-1,0)\times(-1,0)$. Indeed, at $h=4$, the
DBE supports only ``step function'' kinks, 
\beq
\phi(x)=\left\{\begin{array}{cc}
-1 & x<Nh \\
0 & x\geq Nh,\end{array}\right.
\eeq
as may be shown by explicit calculation.

\vspace{0.25cm}
\vbox{
\centerline{\epsfysize=2truein
\epsfbox[90 190 540 630]{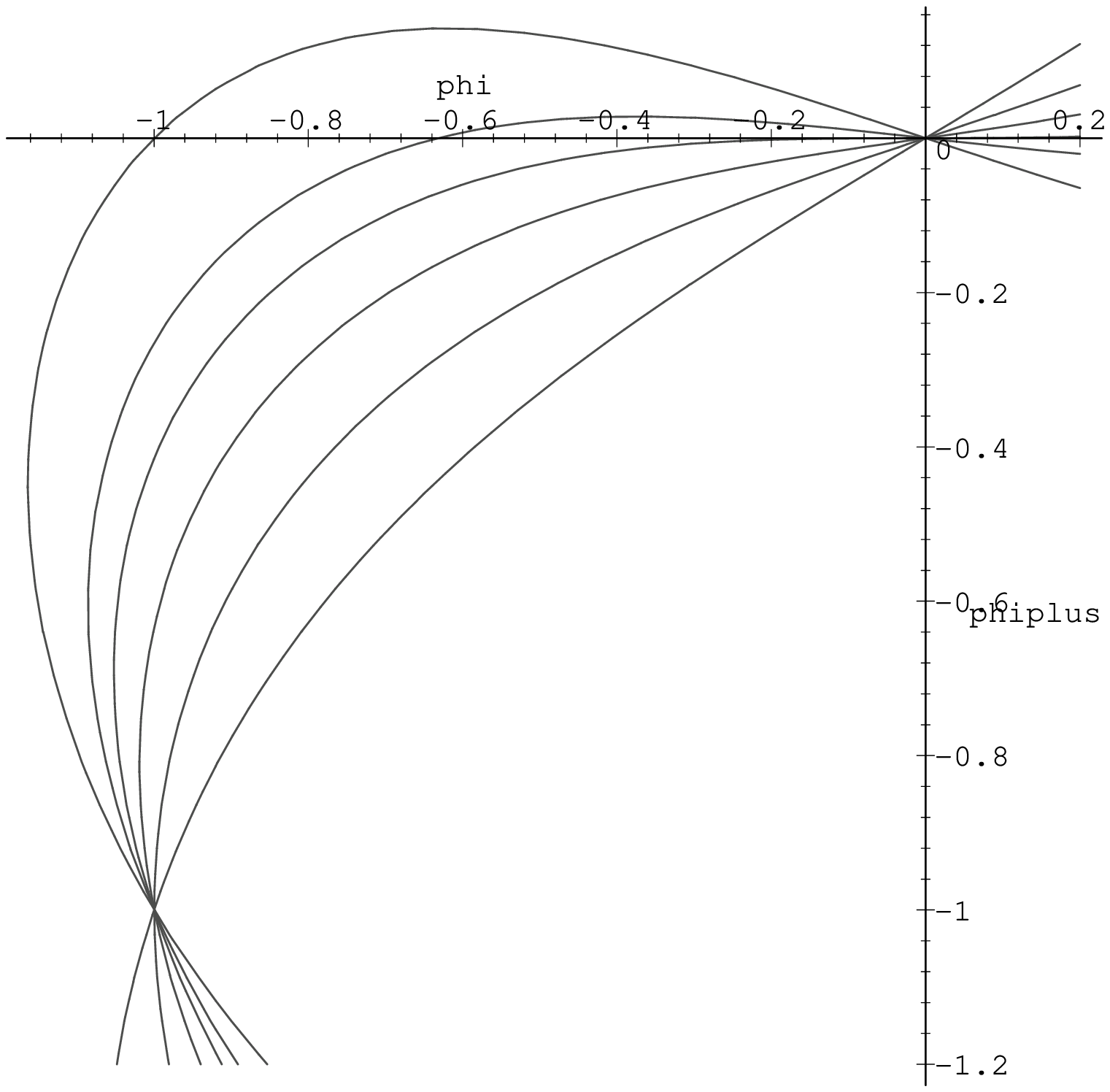}}
\noindent
{\it Figure 4: Zero contours of $\phi_+-\phi-hF(\phi,\phi_+)$ for the values
$h=0.5,1.0,1.5,2.0,2.5$ and $4$, from least to most curved.}
}
\vspace{0.5cm}

Since the DBE (\ref{twbog}) is a cubic polynomial equation, it is possible
in principle to solve for $\phi_+$ in terms of $\phi$ (or {\em vice versa})
and hence generate exact static kink solutions, although it appears that no
closed formula for $\phi(x)$ exists (in particular, a continuum-inspired
ansatz does not seem to work). In general, if $F_c$ is a degree $p$ polynomial,
the DBE is a degree $p$ polynomial equation for $\phi_+(\phi)$, so for 
$p\geq 5$ such an approach would not work. Even for $p=3,4$, implementation
would be complicated. Instead, one may solve the DBE approximately with some
kind of root-finding algorithm. In figure 5 we present a kink solution found
by solving (\ref{twbog}) with $h=0.8$ using the bisection method. Note that
the kink structure is spread over very few lattice sites, so the system is 
deep in the discrete regime.

\vspace{0.25cm}
\vbox{
\centerline{\epsfysize=2truein
\epsfbox[45 390 577 608]{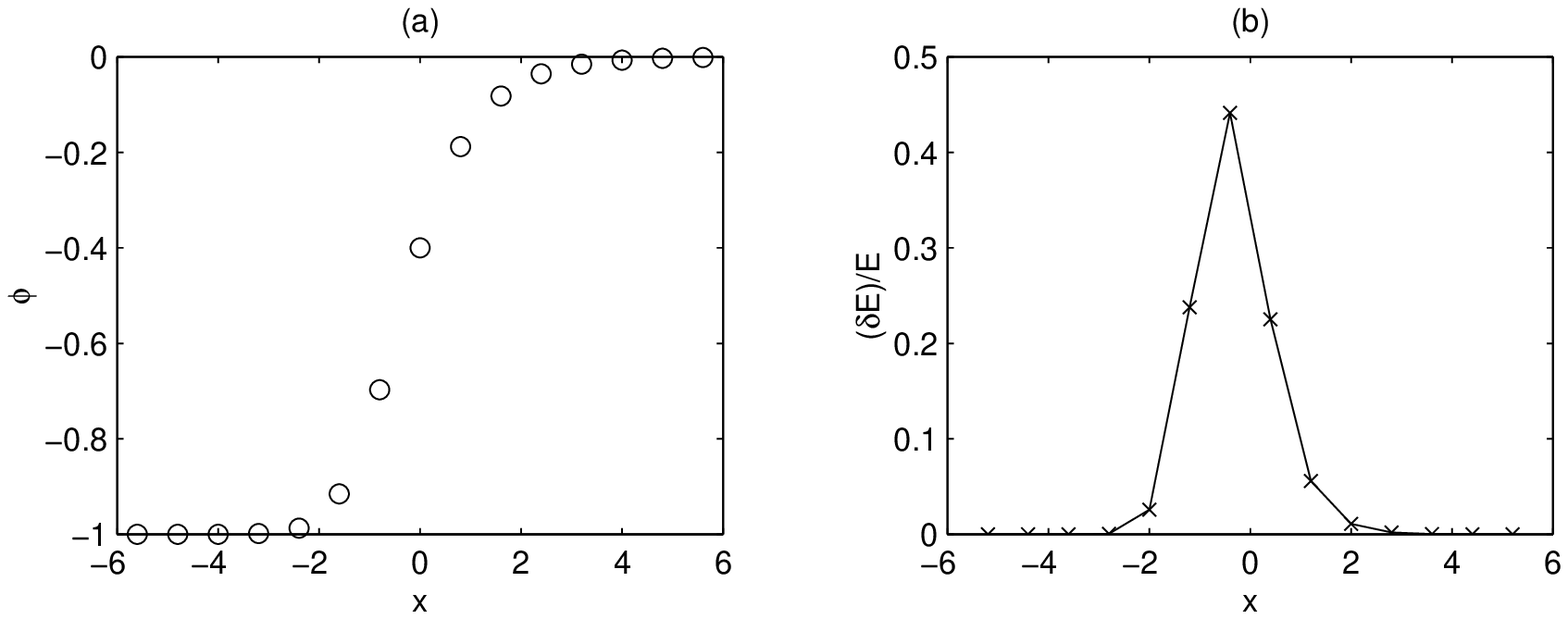}}
\noindent
{\it Figure 5: Example of a static kink solution on a lattice of spacing
$h=0.8$
obtained by approximate solution of the DBE using the bisection method. 
Plot (a) shows the field $\phi$.
Plot (b) shows the energy distribution of the kink. The energy should be thought of as located in the links between pairs of lattice sites. This plot
shows the energy in each link as a fraction of the total energy.}
}
\vspace{0.5cm}

\section{Numerical simulations}
\label{sec:numerics}

So far, we have considered only the static model. In this section we will
discuss numerical simulations of the full, time-dependent discrete field
equation, that is, numerical solutions of the coupled set of ODEs
(\ref{deqm}). In solving these ODEs one must also discretize {\em time} in 
some way. It should be regarded as a strength of our approach that there is 
great freedom in precisely how one does this, since one may choose a method
specifically adapted to the dynamics one seeks to simulate. For example, in
kink-antikink interactions, the solitons usually spend a long time well
separated, then experience a brief, but very violent interaction when they 
finally become close. In this situation a variable time step (rather than
a fixed, rectangular space-time mesh) is clearly sensible. For other problems
it may be desirable to preserve the symplectomorphicity of the phase space
flow (both (\ref{ceqm}) and (\ref{deqm}) are Hamiltonian dynamical systems),
so that a symplectic integrator with fixed time step would be more appropriate.

Since the aim here is to test our discretization's performance, we shall
consider a simple and clear-cut dynamical problem, for which the continuum
behaviour is easily understood: single $(0,1)$ kink motion at constant speed.
The continuum field equation (\ref{ceqm}) is Lorentz invariant, so there
exist Lorentz boosted kinks moving with any speed $v\in[0,1)$. These provide
the initial data, sampled on the discrete lattice $h\Z$, for our simulations.
The lattice spacing was chosen to be $h=0.8$, so that the kink width is 
comparable to $h$. Recall that a major departure of conventional discrete
systems from their continuum counterparts is that solitons suffer radiative
deceleration, and possibly pinning, in the highly discrete regime. An 
important test of the topological discretization, therefore, is to monitor
kink velocity $v(t)$ over simulations of long duration (3000 time units in this
case), to see to what extent radiative deceleration is present. One certainly
expects moving kinks to excite some radiation. For a type $(i,j)$ kink,
this will move with a maximum
group velocity of 
\beq
v_g^{\rm max}:=
\left.\frac{d\omega}{dk}\right|_{\rm max}=1-\frac{h}{2}|F_c'(u_k)|
\eeq
where $k=i$ if $v>0$ or $k=j$ if $v<0$, since phonons are excited behind
the kink, and hence are small oscillations about its trailing vacuum
\cite{thesis}. So for right moving $(0,1)$ kinks, $v_g^{\rm max}=0.2$, while
for left moving $(0,1)$ kinks, $v_g^{\rm max}=0.6$. In either case, one wishes
to avoid radiation being reflected from the fixed boundaries and interfering
with the kink's motion, so the first few sites at either end of the (finite)
lattice are damped. This adds a non-Hamiltonian piece to the flow, so a
symplectic integrator is not appropriate, and we choose to use a fourth order
Runge-Kutta method with fixed time step $\delta t=0.01$. The algorithm, and
the checks made on its accuracy and stability (energy conservation etc.) have
been described previously \cite{spe}.

Simulations were performed for various initial velocities, and in all cases it
was found that kink pinning was eliminated, as one would expect: kinks 
propagate freely, indefinitely. Figure 6 shows plots of $v(t)$ for 
right-moving kinks with initial velocities $v(0)=0.2,0.4,0.6$ and $0.8$.
Figure 7 presents similar graphs for left moving kinks, with $v(0)=-0.2,-0.4,
-0.6$ and $-0.8$. In every case, the upper curve shows $v(t)$ for the 
topological discretization (\ref{deqm}), while the lower curve shows $v(t)$ for
the same initial value problem for the conventional discretization 
(\ref{ndeqm}), solved using the same numerical scheme. 
The thickness of the curves is due to velocity oscillations as the kink
moves from cell to cell through the lattice. These are
 partly an artifact of the way
we define kink position (linear interpolation in this case), and
partly dynamical. In both systems one may regard the effective kink mass as
depending periodically on the kink position, and in the conventional (but not
the topological) discretization, the kink velocity oscillates as it moves up 
and down in the PN potential. As found previously for
sine-Gordon and $\phi^4$ kinks, radiative deceleration is drastically cut 
using the topological discretization.

\vspace{0.25cm}
\vbox{
\centerline{\epsfysize=3truein
\epsfbox{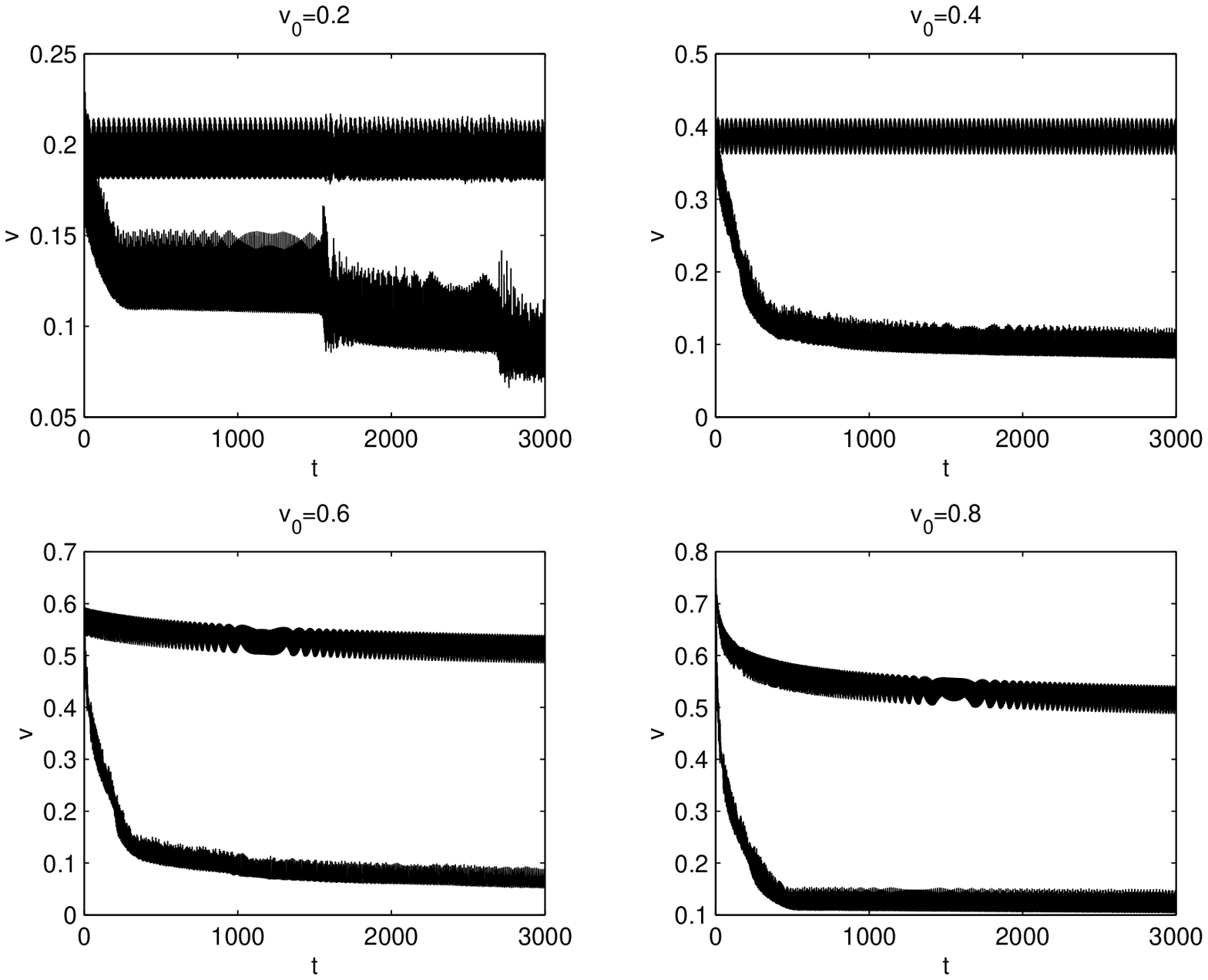}}
\noindent
{\it Figure 6: Comparison of the motion of right-moving
kinks in the
topological discrete twiston system with that in the conventional 
discretization, for various initial velocities. The lattice spacing is $h=0.8$.
In each case, velocity is plotted against time for both systems on the
same graph, the lower curve showing the data for the conventional discrete
system.}
}
\vspace{0.5cm}

\vspace{0.25cm}
\vbox{
\centerline{\epsfysize=3truein
\epsfbox{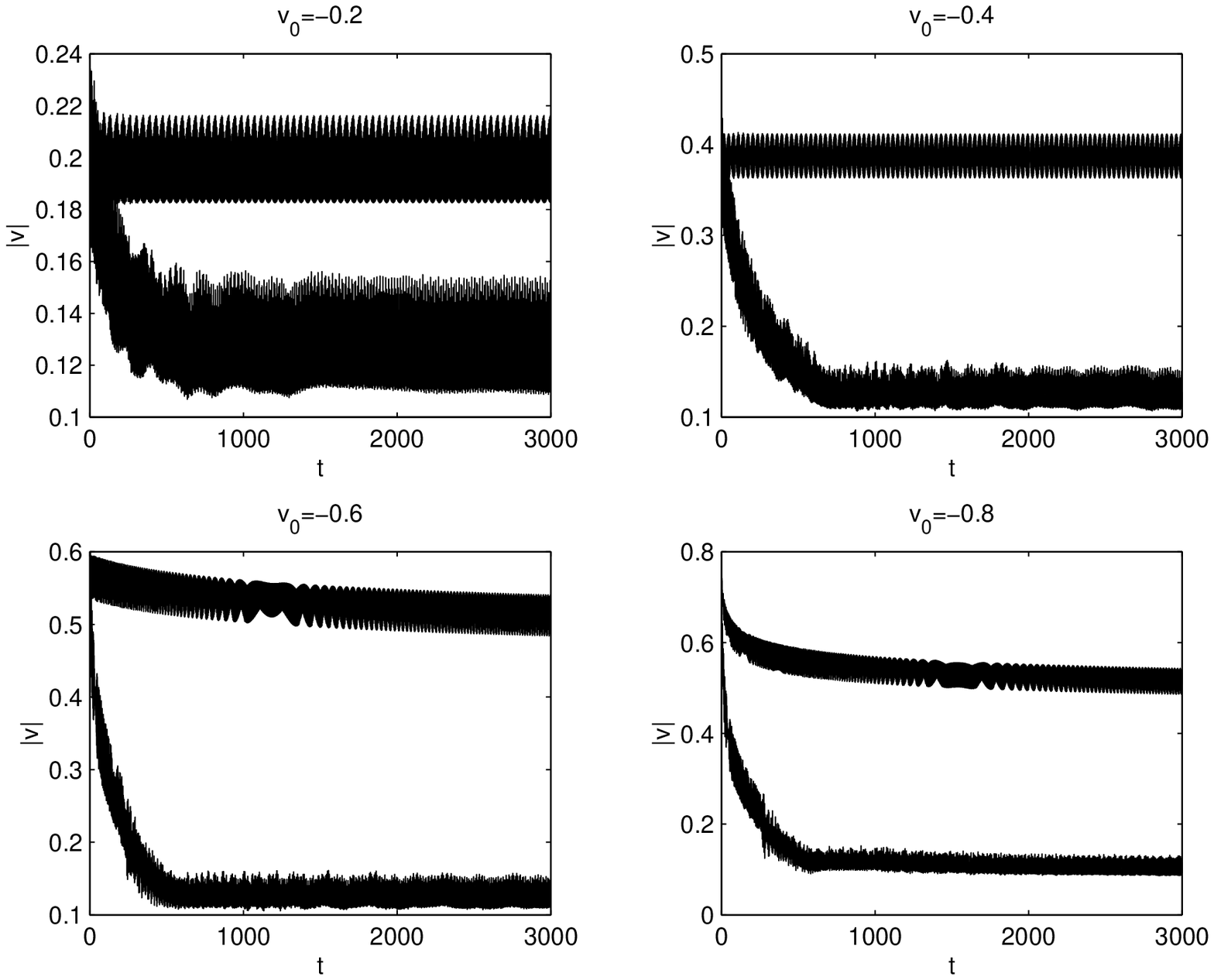}}
\noindent
{\it Figure 6: Comparison of the motion of left-moving kinks in the
topological and conventional discrete twiston systems, with lattice spacing
$h=0.8$.
Speed is plotted against time for both systems on the
same graph, the lower curve showing the data for the conventional discrete
system. As with right-moving kinks, radiative deceleration is much less
pronounced in the topological discretization.}
}
\vspace{0.5cm}

\section{Other methods}
\label{sec:om}

The discretization method outlined in section \ref{sec:topdis} is not the
only way to arrive at discrete systems without a Peierls-Nabarro barrier.
Having performed the construction, one is always free to replace $\Delta\phi$
by $D=X\Delta\phi$ and $F$ by $\wt{F}=F/X$ where $X$ is any function of
$(\phi,\phi_+,h)$ which reduces to unity in the continuum limit,
$h\ra 0$, $\phi_+=\phi+O(h)$. Since $D\wt{F}=\Delta G$, 
the potential energy functional
\beq
E_P=\frac{h}{2}\sum_{x\in h\Z}(D^2+\wt{F}^2)
\eeq
is again subject to the the Bogomol'nyi argument, yielding the same lower 
bound on type $(i,j)$ kink energy, but the DBE is now different:
$D=\pm\wt{F}$. For example, the sine-Gordon model has $F_c=\sin$, and hence
its topological discretization has
\beq
F(\phi,\phi_+)=\left\{\begin{array}{cc}\displaystyle{
-\frac{\cos\phi_+-\cos\phi}{\phi_+-\phi}} & \phi_+\neq\phi \\ & \\
\sin\phi & \phi_+=\phi.\end{array}\right.
\eeq
If one chooses $X={\rm sinc}\, \frac{1}{2}(\phi_+-\phi)$, 
(where ${\rm sinc}\, x:=
x^{-1}\sin x$ when $x\neq 0$ and ${\rm sinc}\, 0:=1$), 
then defining $D=X\Delta\phi$
and $\wt{F}=F/X$ one recovers the original topological discrete sine-Gordon
system \cite{war}, namely
\bea
D&=&\frac{2}{h}\sin\frac{1}{2}(\phi_+-\phi) \nonumber \\
\wt{F}&=&\sin\frac{1}{2}(\phi_++\phi).
\eea 

In general, the kind of analysis carried out in section
\ref{sec:proof} should still work for the new DBE (and certainly does in the
specific case of the sine-Gordon system outlined above), although the
geometric picture of a plane intersecting a surface no longer applies. Using
this trick, one may be able to avoid a piecewise definition of $\wt{F}$ even
when $F_c$ is not polynomial. However, no systematic method for choosing
$X$ suggests itself.

Finally, we wish to describe briefly
a completely different approach due to Flach,
Kladko and Zolotaryuk, called the inverse method. The following is a
somewhat reinterpreted version of the method outlined in \cite{fla}
(the authors of which were primarily concerned with exact {\em moving}\, 
discrete solitons, rather than the Peierls-Nabarro barrier). The idea is
that, given a continuum nonlinear Klein-Gordon model possessing a type $(i,j)$
kink with static profile $\phi_K$, one seeks a one-site substrate potential
$V_h(\phi)$ such that the discrete field equation
\beq
\ddot{\phi}=\frac{\phi_+-2\phi+\phi_-}{h^2}-V'_h(\phi)
\eeq
preserves the one parameter family of static solutions $\phi(nh)=\phi_K(nh-b)$.
Clearly, one needs 
\beq
\label{*}
V'_h(\phi_K(z))=\frac{1}{h^2}[\phi_K(z+h)-2\phi_K(z)+\phi_K(z-h)]
\eeq
for all $z\in\R$. This uniquely determines $V'_h:I_{ij}\ra\R$ by monotonicity
of $\phi_K$. Note that, by construction, in the limit $h\ra 0$, the right hand
side of (\ref{*}) becomes $\phi_K''$ so, given that $\phi_K$ satisfies
the static continuum field equation (\ref{ceqm}), 
$V_h'(\phi)$ must have the correct
continuum limit ($\lim_{h\ra 0} V'_h=F_cF'_c$). 
Since this is a continuous family
of static solutions, they must all have the same energy. Indeed, equation
(\ref{*}) may be interpreted as the condition 
\beq
\frac{d\, }{db}E_P[\phi_K(nh-b)]=0,
\eeq
where the potential energy is
\beq
E_P[\phi]=h\sum_{x\in h\Z}\left[\frac{1}{2}(\Delta\phi)^2+V_h(\phi)\right].
\eeq
Hence there is no PN barrier resisting the propagation of kinks in this 
discrete system.
If $\phi_K$ is analytic, $V'_h(\phi)$ can be written as a power
series in $\phi$, although there may exist no closed formula for $V'_h$. This
may be a significant drawback of the inverse method, in comparison with
the topological discretization approach, which works very explicitly, at least
for all polynomial $F_c$.

In some special cases, $V'_h$ can be obtained in closed form: one example is
$\phi^4$ theory, where $\phi_K=\tanh$ and an explicit formula for $V_h$ has
been known for some time \cite{sch}. As a new example, one could consider
the continuum twiston model defined in section \ref{sec:twist}, whose $(2,1)$
kink has profile $\phi_K(x)=[\frac{1}{2}(1-\tanh x)]^\frac{1}{2}$. 
In this case,
$\phi_K(z\pm h)$ can be written in terms of $\phi_K(z)$ using hyperbolic
trigonometric identities, so one finds,
\beq
\label{imtw}
V'_h(\phi)=\frac{\phi}{h^2}\left\{\sqrt{\frac{1-\tanh h}{1+(1-2\phi^2)\tanh h}}
+\sqrt{\frac{1+\tanh h}{1-(1-2\phi^2)\tanh h}}-2\right\},
\eeq
which has $V'_h(\pm 1)=V'_h(0)=0$, as expected. It is unclear how ``special''
$F_c$ must be in order for the $V'_h$ obtained by this inverse method to be
expressible in closed form -- perhaps this happens whenever $F_c$ is 
polynomial. If so, one 
would expect the expressions involved to become extremely
complicated as the degree of $F_c$ grows. The inverse method also works
explicitly when applied to the sine-Gordon kink profile, $\phi_K(x)=
2\tan^{-1}e^x$, so generating yet another discrete sine-Gordon system with
no PN barrier (see also \cite{zak}). In this case,
\beq
\label{imsg}
V'_h(\phi)=\frac{2}{h^2}\left[\tan^{-1}\left(e^h\tan\frac{\phi}{2}\right)
+\tan^{-1}\left(e^{-h}\tan\frac{\phi}{2}\right)-\phi\right].
\eeq
Equations (\ref{imtw}) and (\ref{imsg}) can be integrated to give explicit
formulae for $V_h$, and hence $E_P$, for these models, although the formulae
are rather complicated. This is little practical disadvantage, since only
$V'_h$ appears in the discrete field equation.
The strength of this method is that
one has an explicit formula for the family of static discrete kinks, instead
of an abstract existence theorem. It certainly merits further investigation.

\vs\vs
\noindent
{\bf Acknowledgments}
\vs
\newline
The author wishes to thank S. Flach and Y. Zolotaryuk for their patient
explanation of the inverse method.

\end{document}